\documentclass[sigplan,nonacm]{acmart}
\usepackage{booktabs}
\usepackage{colortbl}
\usepackage{multirow}
\usepackage{xspace}

\graphicspath{{figures/}}

\newcommand{\sysname}{ServeTwin\xspace}
\newcommand{\etgen}{iSTAGE\xspace}

\begin{document}

\title{\sysname:
A Benchmark-Validated Simulator for Distributed LLM Architecture Exploration
}

\author[Park et al.]{\texorpdfstring{%
Sungjoon Park\textsuperscript{*}\quad
Changue Jung\textsuperscript{*}\quad
Kyungno Joo\quad
Mincheol Kang\\
Jaehyung Ahn\quad Sehwan Lee\quad Sangjoon Kim%
}{Sungjoon Park; Changue Jung; Kyungno Joo; Mincheol Kang; Jaehyung Ahn; Sehwan Lee; Sangjoon Kim}}
\email{{stdio.park,changue.jung,kyungno.joo,minc.kang,jh91.ahn,sehwan.b.lee,sangjoon0919}@samsung.com}
\affiliation{\institution{Samsung Advanced Institute of Technology}\city{Suwon-si}\country{Republic of Korea}}

\begin{abstract}

Evaluating distributed LLM serving designs on physical clusters is costly. Yet existing simulators provide only subsets of the capabilities needed for realistic design exploration: stateful closed-loop execution, timing prediction without profiling target hardware, and direct execution of unmodified serving benchmarks. We present \sysname, a closed-loop simulator that couples specification-driven analytical timing with a stateful serving loop. This coupling captures feedback among request completion, scheduling, queue state, and KV-cache evolution, including prefill--decode disaggregation and multi-turn agent workloads.

\sysname avoids target-hardware operator profiling through iSTAGE, an analytical trace generator that derives per-iteration traces from model
and engine specifications while representing ragged batches and discrete engine mechanisms such as CUDA-graph batch padding. It further decomposes execution time into component-owned throughput, scheduler, and runtime costs. This ownership lets unaffected parameters transfer across platforms and confines recalibration to changed hardware or software components. \sysname implements a vLLM-compatible interface and runs unmodified serving benchmarks. Against real deployments, it reproduces InferenceX's steady-state throughput--interactivity frontier with a 3.6\% mean error and predicts LMBenchmark's multi-turn performance with a 9.9\% error while tracking KV-cache evolution. Pre-silicon sweeps reveal that the preferred HBM bandwidth--capacity tradeoff reverses across workload states and that increasing concurrency shifts the bottleneck from memory bandwidth to scheduler and runtime overheads. Together, these capabilities enable practical exploration of distributed LLM serving systems before target hardware is available.

\end{abstract}

\maketitle

\begingroup
\renewcommand{\thefootnote}{*}
\footnotetext{The first two authors contributed equally to this work.}
\endgroup

\raggedbottom

\begin{figure*}[t]
\begin{minipage}[t]{0.48\textwidth}
\centering
\includegraphics[width=\linewidth]{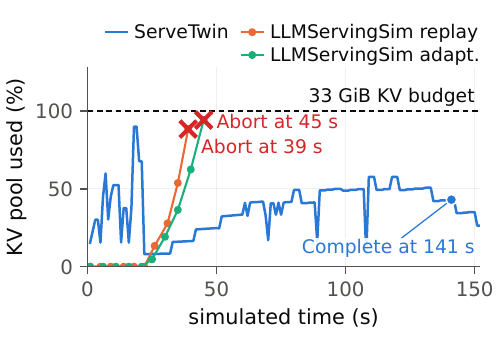}
\captionof{figure}{\textbf{Capacity failure under a deployment-equivalent KV budget.} LLMServingSim aborts the multi-turn workload without a staged-KV waiting state. The 33\,GiB KV budget is derived from an 80\,GB H100.}
\label{fig:state-failure}
\end{minipage}
\hfill
\begin{minipage}[t]{0.48\textwidth}
\centering
\includegraphics[width=\linewidth]{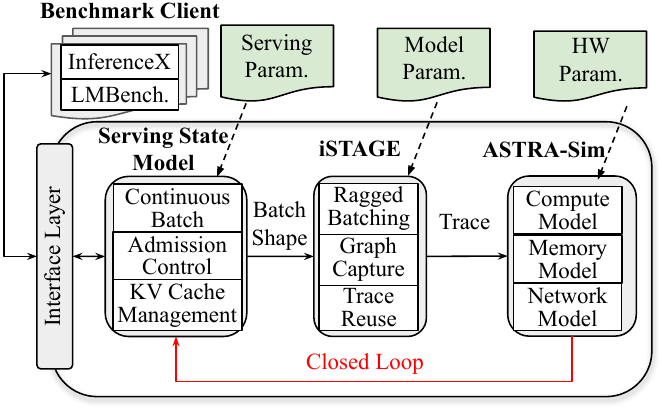}
\captionof{figure}{\textbf{The closed-loop simulation architecture of \sysname{}.} Simulated execution times from \etgen{} and ASTRA-sim dynamically dictate subsequent scheduling decisions in a continuous feedback loop.}
\label{fig:overview}
\end{minipage}
\end{figure*}

\section{Introduction}
\label{sec:intro}

Large Language Model (LLM) serving extends far beyond accelerating isolated tensor operations; it has evolved into a highly dynamic systems engineering challenge. Fundamentally, the autoregressive nature of LLMs dictates a strict two-phase execution for every request: a prefill phase that processes the prompt to generate a KV cache, followed by an iterative decode phase. While modern engines like vLLM~\cite{vllm} and SGLang~\cite{sglang} employ continuous batching~\cite{orca} to dynamically mix these phases and maximize throughput, the differing compute profiles of prefill and decode have driven a recent architectural shift. Today, modern deployments physically disaggregate these two phases onto separate workers, coupling them via KV cache transfer~\cite{dynamo}. This architectural complexity is further amplified by agentic workloads. Unlike simple pass-through queries, agentic clients introduce a strict multi-turn dependency, issuing subsequent requests only after receiving the previous response. This shift transforms LLM serving into a highly interactive, stateful system loop, demanding sophisticated evaluation. 

\begin{table}[t]
\caption{\textbf{Comparison of prior tools against ServeTwin's key simulation capabilities.}
ServeTwin uniquely combines stateful closed-loop execution, profiling-free timing prediction,
and compatibility with unmodified serving benchmarks.
($\bigcirc$~full, $\triangle$~partial, $\times$~absent.)}
\label{tab:gap}
\footnotesize
\centering
\setlength{\tabcolsep}{3.5pt}
\begin{tabular}{@{}lccc@{}}
\toprule & \shortstack{Stateful\\serving loop}
 & \shortstack{Profiling\\free}
 & \shortstack{Unmodified\\benchmarks} \\
\midrule
LLMServingSim 2.0~\cite{llmservingsim2} & $\triangle$ & $\times$ & $\times$ \\
Vidur~\cite{vidur} & $\triangle$ & $\times$ & $\times$ \\
AgentServeSim~\cite{agentservesim} & $\triangle$ & $\times$ & $\triangle$ \\
GenZ~\cite{genz} / LLMCompass~\cite{llmcompass} & $\times$ & $\bigcirc$ & $\times$ \\

\rowcolor{black!8}
\textbf{\sysname} & $\bigcirc$ & $\bigcirc$ & $\bigcirc$ \\
\bottomrule
\end{tabular}
\end{table}

Production metrics like TTFT and queue transients emerge from a continuous feedback loop: serving state dictates each iteration's batch, while hardware execution time updates the subsequent state (e.g., queue depth and KV-cache occupancy). Consequently, faster hardware inherently alters the batches themselves, rendering fixed-batch models inadequate. While evaluating architectural improvements under these dynamic conditions is critical, the prohibitive scale and cost of LLM infrastructure drives an urgent demand for high-fidelity simulators that capture this end-to-end dynamic environment. Rapidly evolving agentic workloads compound that demand.

Existing simulators fail to capture this continuous loop. Analytical models (e.g., GenZ~\cite{genz}, LLMCompass~\cite{llmcompass}) evaluate static workloads without queueing feedback, while serving simulators (e.g., LLMServingSim~\cite{llmservingsim}, Vidur~\cite{vidur}) reproduce scheduling but rely on rigid physical hardware profiles. Crucially, current tools fundamentally fail at dynamic state management, particularly in prefill/decode disaggregation. For example, LLMServingSim treats prefill/decode disaggregation as a simple pipeline cut. It lacks a way for the prefill worker to temporarily hold the KV cache if the decode worker is busy. As Figure~\ref{fig:state-failure} shows, this creates a fundamental simulation failure, not just a timing inaccuracy. Multi-turn workloads simply crash because requests attempt to enter the decode phase without any queue to safely wait in.

\sysname{} reconnects this severed feedback loop and resolves these state failures, as illustrated in Figure~\ref{fig:overview}. Running unmodified multi-turn benchmarks, it completes execution under deployment-equivalent KV budgets by properly staging the waiting path. Each iteration, it constructs a batch mirroring production LLM engines (e.g., vLLM, SGLang), strictly enforcing admission control, cross-worker KV transfers, and multi-turn states. To process this without empirical profiles, we introduce iSTAGE—extending STAGE~\cite{stage} to translate the batch into a symbolic trace, calculating latency directly from hardware specifications.
Feeding this precise timing back to advance the system state naturally dictates the next batch composition.

Integrating an analytical model within a dynamic serving loop poses profound challenges. First, serving batches are inherently ragged; naïvely padding them to maximum dimensions for analytical graphs overestimates execution time by up to 146$\times$. Second, theoretical FLOPs and memory bandwidth cannot capture discrete runtime mechanisms (e.g., KV page walks, kernel launches) that heavily skew step latency. Finally, continuous end-to-end simulation demands exceptionally lightweight trace generation. iSTAGE overcomes these hurdles through (1) statistics-preserving batch binding, (2) first-order modeling of discrete engine mechanisms, and (3) aggressive trace reuse.

We validate \sysname{} across two distinct evaluation paradigms. For steady-state performance, InferenceX~\cite{inferencex} evaluates system throughput and interactivity frontiers. Because it largely operates without the unpredictability of prefix-cache hits, \sysname{} easily tracks physical performance with just a 3.6\% mean error. In contrast, LMBenchmark~\cite{lmbench} introduces the chaotic dynamics of multi-turn serving. In this environment, real-time prefix-cache hits constantly shift the execution timing balance between prefill and decode (P/D) phases, dynamically cascading into admission control and queueing variations. Despite this extreme runtime complexity where prior methods simply crash, \sysname{} successfully captures these transient dynamics within a 10\% error margin.

This paper makes the following contributions:

\begin{enumerate}
\item \textbf{Stateful Serving Model:} A dynamic serving model that explicitly tracks request scheduling, multi-turn sessions, and KV-cache lifecycles, featuring occupancy-coupled admission for prefill/decode disaggregation rather than simplistic fixed-delay handoffs (\S\ref{sec:fidelity}--\S\ref{sec:sessions}).

\item \textbf{Spec-Driven Execution (\etgen):} An analytical trace generator that derives execution timings directly from hardware and model specifications, eliminating the need for empirical profiles while remaining lightweight enough to operate inside the serving loop (\S\ref{sec:etgen}).

\item \textbf{Component-Owned Cost Decomposition:} A mechanism-attributed cost ledger that separates specification-driven throughput from host-control and runtime-mechanism costs. Each term family is assigned an observable ablation footprint and an owning component, enabling selective recalibration only when that component changes instead of global overhead fitting or per-point retuning (\S\ref{sec:ledger}, \S\ref{sec:validation}--\S\ref{sec:eval-transfer}).

\item \textbf{Benchmark-Validated Architectural Exploration:} Validation against unmodified public benchmarks across hardware generations, serving stacks, and deployment scales, followed by spec-driven hardware sweeps that reveal a workload-dependent trade-off between HBM bandwidth and capacity and a bottleneck shift from memory physics toward runtime control overheads (\S\ref{sec:eval-gates}--\S\ref{sec:casestudy}).
\end{enumerate}

\section{Background and Motivation}
\label{sec:background}

\subsection{LLM Serving as a Stateful System}
\label{sec:anatomy}
LLM serving entails two distinct phases: prefill (processing prompts to generate a KV cache) and decode (iteratively generating tokens, expanding the cache). Modern engines dynamically reconstruct batches every iteration via continuous batching and chunked prefill, while leveraging prefix caching for history reuse. Crucially, execution is fundamentally state-dependent. A request is admitted only if sufficient KV memory exists; otherwise, it queues or preempts. Because discrete runtime mechanisms often reshape physical execution (\S\ref{sec:mechanisms}), simulators must explicitly track the dynamic system state (e.g., KV occupancy) that governs batch composition, rather than merely counting requests.

\subsection{Prefill/Decode Disaggregation and KV Transfer}
\label{sec:pd}
Disaggregated architectures~\cite{splitwise,distserve,mooncake} isolate prefill and decode workers to prevent long prompts from stalling token generation. This necessitates transferring the KV cache between workers. However, this handoff is not a static network delay. Production systems (e.g., Dynamo~\cite{dynamo} with NIXL~\cite{nixl}) implement a pull mechanism where decode workers fetch KV blocks only when they have sufficient free capacity. If decode memory is full, the cache remains staged on the prefill worker, creating upstream backpressure. Accurate simulation requires tracking this \emph{occupancy-coupled admission} and the resulting capacity-driven dependencies (\S\ref{sec:lifecycle}).

\subsection{Benchmarks Used in This Paper}
\label{sec:benchmarks}
We validate \sysname{} using two distinct benchmarks.

\textbf{InferenceX}~\cite{inferencex} maps the steady-state balance between throughput and interactivity. We validate our simulator against the public hardware performance data published online by SemiAnalysis.

\textbf{LMBenchmark}~\cite{lmbench} captures dynamic, multi-turn behaviors like admission pressure and transient delays. We validated this by deploying an in-house DGX H100 with NVIDIA Dynamo and vLLM, directly matching \sysname{}'s behavior with physical telemetry data extracted via Prometheus and Grafana.

\subsection{Existing Simulation Approaches}
\label{sec:existing}

\textbf{Serving Simulators.} Tools like LLMServingSim~\cite{llmservingsim,llmservingsim2}, Vidur~\cite{vidur}, and AgentServeSim~\cite{agentservesim} model continuous batching and scheduling but fundamentally rely on hardware-profiled latencies. Furthermore, they omit critical stateful semantics: LLMServingSim relies on open-loop trace replay that breaks multi-turn session dependencies, and models prefill-to-decode disaggregation as an unconditional pipeline cut lacking staged-KV backpressure. Vidur and AgentServeSim similarly lack occupancy-coupled KV lifecycles.

\textbf{Analytical Models.} GenZ~\cite{genz} and LLMCompass~\cite{llmcompass} estimate execution timings directly from hardware specifications, enabling architectural exploration. However, they evaluate static workloads offline (i.e., snapshots), completely lacking the dynamic closed-loop scheduling required to iteratively update batch compositions.

\textbf{STAGE.} STAGE~\cite{stage} synthesizes symbolic tensor graphs for distributed LLM workloads and includes an inference case study. Its abstraction, however, models prescribed workload graphs rather than a stateful serving loop: it does not represent KV-cache lifecycles, queueing feedback, or the scheduler-generated ragged batches that change at every iteration. Adapting it for serving therefore requires the serving-specific mechanisms described in \S\ref{sec:etgen}.

\subsection{Why Existing Serving Simulation Is Insufficient}
\label{sec:motivation}

LLMServingSim represents prefill/decode disaggregation as a pipeline
cut and has no state in which a completed prefill request can retain
its KV on the producer while waiting for decode-side capacity.
Figure~\ref{fig:state-failure} gives a motivating counterexample.
\sysname{} runs the unmodified multi-turn benchmark and completes
under the deployment-equivalent KV budget, while both LLMServingSim
encodings abort when they attempt to allocate the decode-side KV
without a staged waiting path. This is a semantic
failure rather than an execution-time calibration error: all methods
use the same calibrated compute profile.

\section{\sysname{} Design}
\label{sec:overview}

\subsection{Overview}
\label{sec:simserver}

Figure~\ref{fig:overview} illustrates \sysname{}'s closed-loop architecture. Unmodified clients interact via a lightweight wall-clock-to-simulated-time interface. Each iteration executes four stages: (1) The stateful serving loop admits and batches requests under dynamic memory and token budgets, tracking KV lifecycles. (2) iSTAGE binds the batch shape to a symbolic tensor graph, emitting a Chakra execution trace~\cite{chakra}. (3) ASTRA-sim~\cite{astrasim} evaluates the trace; employing a structural network simulator rather than closed-form equations captures topology-dependent collective communication overheads. (4) The resulting execution time advances the simulated clock and updates system states to drive subsequent batching.

\subsection{Serving-State Simulation}
\label{sec:fidelity}

While the baseline scheduling logic builds upon LLMServingSim~\cite{llmservingsim,llmservingsim2}, we redesign its internal state management to achieve strict closed-loop fidelity. Through this redesign, \sysname{} models production systems like vLLM and SGLang, supporting continuous batching, chunked prefill, radix-tree prefix caching, and capacity-driven preemption. To track these dynamics, we introduce a KV lifecycle registry featuring Dynamo-inspired pull-based handoffs (\S\ref{sec:pd}), occupancy-coupled admission, tiered memory, and a dedicated multi-turn session layer. Finally, iSTAGE and ASTRA-sim replace the legacy profile-based backend, enabling fully spec-driven simulation.

\begin{figure}[t]
\centering
\includegraphics[width=\columnwidth]{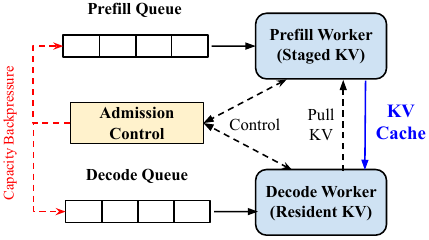}
\caption{\textbf{Capacity-driven KV handoff in disaggregated serving.} Decode-side capacity regulates KV transfers, naturally backpressuring the prefill stage when the pool is full.}
\label{fig:lifecycle}
\end{figure}

\subsection{Disaggregated KV Lifecycle}
\label{sec:lifecycle}

In \sysname, a request's KV cache transitions through an explicit state machine (Figure~\ref{fig:lifecycle}), from initial allocation to eventual eviction or demotion. Unlike a simplistic pipeline cut, this lifecycle enforces two critical couplings. First, admission is \emph{occupancy-coupled}: decode-side memory saturation blocks admission, naturally inducing backpressure on the prefill queue. Second, KV transfer is strictly \emph{consumer-driven}: staged KV awaits an explicit pull from the decode worker (modeling NIXL-style handoffs~\cite{nixl}). Decoding commences only upon full KV arrival or suffix recomputation, supporting both full and layer-wise transfers.

Implementation components include a prefill-side registry with heartbeat leases, a decode-side wait queue, bandwidth-constrained transfer tracking, radix-tree-aware delta pulls, and a recompute fallback. By explicitly modeling these mechanisms, \sysname{} captures crucial second-order dynamics—such as localized queue buildups, network backlogs, and KV-induced admission stalls—that unconditional pipeline models obscure.

\subsection{Multi-turn Sessions as Simulator State}
\label{sec:sessions}

Unlike traditional trace-driven approaches that statically unroll conversations into independent requests, \sysname{} tracks multi-turn sessions as internal simulator state. A session issues a subsequent turn only after the preceding response is fully generated, reflecting strict inter-request dependencies and optional tool-execution delays. Because context accumulates continuously across rounds, the integrated prefix cache (\S\ref{sec:fidelity}) translates shared history into cache hits and minimal delta transfers.

Furthermore, \sysname{} supports tiered KV memory management: blocks evicted from HBM can be demoted to host memory and promoted back upon subsequent hits, with all data movement strictly constrained by the host-to-device interconnect bandwidth. Additionally, speculative decoding is analytically incorporated via a parameterized per-step acceptance length. Collectively, these state-tracking mechanisms capture the workload semantics inherent to the multi-round agentic scenarios evaluated by LMBenchmark.

\begin{figure}[t]
\centering
\includegraphics[width=\columnwidth]{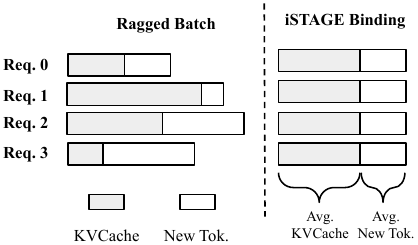}\par
\smallskip
{\small (a)}\par
\medskip
\includegraphics[width=\columnwidth]{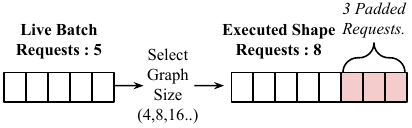}\par
\smallskip
{\small (b)}\par
\caption{\textbf{Mapping serving batches to execution shapes.} (a) \etgen{} binds ragged batches to statistics-preserving rectangles. (b) Decode batches are padded to discrete CUDA-graph capture sizes.}
\label{fig:istage}
\end{figure}

\subsection{Spec-Driven Execution Modeling with iSTAGE}
\label{sec:etgen}

While STAGE~\cite{stage} provides foundational spec-driven graph synthesis for distributed LLM workloads, it does not model the dynamic serving state or KV-aware ragged batches produced by a serving scheduler. To bridge this gap, we develop \textbf{iSTAGE (Inference STAGE)}, which extends and retargets this symbolic execution modeling specifically for serving. This enables accurate, profile-free timing evaluation based purely on hardware specifications. iSTAGE achieves modularity via configuration files rather than hardcoded backends. As summarized in Table~\ref{tab:features}, its templates support a wide spectrum of modern architectural features. This breadth is essential for architectural exploration, as varying model structures fundamentally shift system bottlenecks across compute, bandwidth, memory capacity, and communication.

However, integrating these models into an iterative serving loop introduces three unique challenges absent in training simulations: (1) modeling irregular (ragged) batches without padding-induced cost distortions; (2) capturing discrete engine mechanisms that alter latency independently of theoretical FLOPs/bytes; and (3) ensuring trace generation is lightweight enough for per-iteration execution. As illustrated in Figure~\ref{fig:istage}, iSTAGE resolves these by replacing naïve rectangular schemas with statistics-preserving batch binding and applying capture-size padding to reflect actual engine discreteness.

\paragraph{Ragged Batches.}
\label{sec:meanfield}

Symbolic execution graphs typically require rectangular inputs, but LLM serving batches are inherently ragged, comprising requests with varying new tokens ($q_i$) and KV lengths ($kv_i$). Padding to the maximum lengths (max-folding) substantially overestimates compute costs, especially in mixed prefill--decode batches.

To resolve this without losing analytical efficiency, we observe that the underlying cost model depends only on aggregate statistics: total tokens ($\sum q_i$ for dense GEMMs), attention FLOPs ($\sum q_i kv_i$), and KV footprint ($\sum kv_i$). Therefore, iSTAGE binds the graph dimensions to statistics-preserving averages: $\mathrm{Seq}=\frac{1}{n}\sum q_i$ and $\mathrm{KVSeq}=\sum q_i kv_i / \sum q_i$ as shown in Figure~\ref{fig:istage}a. This guarantees exact token and attention FLOP counts, while precisely matching the KV footprint during pure-decode phases ($q_i{=}1$).

Against exact ragged computation, iSTAGE yields a median step-time error of $3.7\%$ (p95 $11\%$) on mixed prefill--decode batches, versus $268\%$ for max-folding. Across all iterations, including pure decode, iSTAGE's median error is $1.1\%$.

\paragraph{Engine Discreteness.}
\label{sec:mechanisms}

Beyond batch representation, accurate simulation must account for runtime engine interventions that alter physical workloads. For instance, execution engines routinely pad decode batches to match pre-compiled CUDA-graph capture sizes. This padding constitutes real computational work; in MoE models, it artificially increases the number of distinct experts fetched from HBM, changing actual latency independently of the logical token count (Figure~\ref{fig:istage}b). To capture this, iSTAGE explicitly models the engine's discrete behaviors and capture lists as first-order effects (Table~\ref{tab:features}). Consequently, in \sysname{}, \emph{performance curve shapes emerge inherently from structural mechanisms, while fitted constants merely calibrate baseline levels.}

\paragraph{Keeping Trace Generation Affordable.}
\label{sec:speed}

Serving simulations run tens of thousands of iterations, making on-the-fly trace generation a major performance bottleneck. This overhead historically forced analytical models to rely on offline, single-shot evaluations. To overcome this, iSTAGE avoids redundant work by aggressively reusing previously generated traces. It caches recurring batch shapes, collapses identical model layers (layer folding), and skips through repeated decode steps. This reuse does not sacrifice accuracy; layer folding, for example, remains perfectly exact by explicitly re-adding the per-collective network latencies (\S\ref{sec:ledger}). Ultimately, these caching optimizations (\S\ref{sec:eval-cost}) make trace generation lightweight enough to run efficiently inside the closed loop on a single CPU, without GPU acceleration.

\subsection{Component-Driven Latency Modeling}
\label{sec:ledger}

Hardware specifications easily provide baseline times for compute and memory traffic. However, real-world execution includes complex overheads like host scheduling, KV cache traversal, and network startups. If a simulator lumps all these delays into a single, generic "overhead" constant, it cannot accurately predict performance when hardware or software configurations change.

To address this, \sysname{} breaks down every overhead by its actual root cause—a principle we call \emph{component ownership}. Every delay in our model must map to a specific system operation, scale based on clear variables (like request count or context length), and remain unchanged unless that specific operation changes. Consequently, the total time for each iteration is the sum of three distinct components:
\begin{equation}
\label{eq:step}
t_{\mathrm{step}} = t_{\mathrm{throughput}} + t_{\mathrm{scheduler}} + t_{\mathrm{runtime}}
\end{equation}

\begin{table}[t]
\caption{ \textbf{\etgen{} coverage.} Measured evaluation covers DeepSeek-R1 and Qwen3; additional model families exercise the same analytical templates in simulation.
}
\label{tab:features}
\footnotesize
\centering
\begin{tabular}{@{}p{0.24\columnwidth}p{0.68\columnwidth}@{}}
\toprule
Category & Representative support \\
\midrule
Models & DeepSeek-R1, Qwen3 (measured); Llama, DeepSeek-V4, Kimi K2/K3 (template coverage) \\
\addlinespace[2pt]
Mode & Training, Inference \\
Structure & Dense, MoE \\
Attention & GQA, MLA, CSA, Linear, Hybrid attention \\
Features & P/D Disagg., Chunked prefill, Ragged mixed batches, KV delta pooling \\
\bottomrule
\end{tabular}
\end{table}

\textbf{Throughput Cost ($t_{\mathrm{throughput}}$):} The fundamental time spent on GPU computation, memory access, and network transfer. For prefill, we use a compute utilization curve that scales with the number of tokens. For decode, execution is typically limited by memory bandwidth; tensor-core tile quantization additionally bounds compute efficiency at small batches.

\textbf{Scheduler Cost ($t_{\mathrm{scheduler}}$):} This category includes per-step and per-request CPU overheads and additional GPU work induced by CUDA-graph batch padding. Although padding increases GPU computation and expert-weight traffic, we attribute this additional cost to the Scheduler because it arises from the scheduler's batch-size decision.

\textbf{Runtime Cost ($t_{\mathrm{runtime}}$):} Engine-specific runtime delays. This explicitly includes the time to search the KV cache (which grows with context length), GPU kernel launch delays (which become noticeable at small batch sizes), and the setup time for collective communications. Under data-parallel attention (DPA), ranks finish their work at different times; the DPA residual captures both the resulting wait for slower ranks and the additional synchronization-barrier overhead. The Runtime category also includes P$\to$D pacing, which limits the handoff rate rather than adding to $t_{\mathrm{runtime}}$ in Equation~\ref{eq:step}.

By structurally isolating these costs, \sysname{} avoids arbitrary curve fitting. Our parameters represent real system mechanics, allowing the model to generalize accurately across diverse architectural designs.

\begin{table}[t]
\caption{\textbf{Cost ledger.} Cost parameters are updated only when the corresponding hardware or software component changes.}
\label{tab:ledger}
\footnotesize
\centering
\setlength{\tabcolsep}{3pt}
\renewcommand{\arraystretch}{1.08}
\begin{tabular}{@{}p{0.18\columnwidth}p{0.28\columnwidth}p{0.46\columnwidth}@{}}
\toprule
Category & Cost factor & Architectural source \\
\midrule
\centering Throughput\newline \makebox[\linewidth][c]{\textbf{(T)}} & GEMM utilization & Tensor core efficiency \\
\midrule
\multirow{2}{=}{\centering Scheduler\newline \makebox[\linewidth][c]{\textbf{(S)}}} & Host control & CPU stack overhead \\
& Capture behavior & Graph padding penalty \\
\midrule
\multirow{5}{=}{\centering Runtime\newline \makebox[\linewidth][c]{\textbf{(R)}}} & KV traversal & Attention page-walk \\
& Launch exposure & Kernel submission gap \\
& Collective residual & Comm. setup \& barrier \\
& DPA residual & Workload imbalance \& barrier \\
& P$\to$D pacing & KV transport setup \\
\bottomrule
\end{tabular}
\end{table}

Crucially, these mechanism costs do not overlap with the network simulator. While ASTRA-sim handles the actual data transfer time, our calibrated terms isolate only the auxiliary startup and synchronization delays; the two components divide the work clearly without double-counting. Table~\ref{tab:ledger} summarizes these costs and their system owners. To prevent these terms from becoming unjustified fitting parameters, \sysname{} enforces three strict rules:

\begin{enumerate}
    \item \textbf{Use specifications directly:} Peak compute, memory bandwidth, and engine capture lists are drawn strictly from hardware and software specifications, never from curve fitting.
    \item \textbf{Ground terms in observation:} We introduce overhead terms exclusively when warranted by empirical scaling trends. For instance, a cost profile scaling with context length isolates KV-cache traversal, whereas an overhead persisting across hardware transitions signifies software-stack overheads.
    \item \textbf{Fix and reuse values:} Calibrated parameters are frozen and consistently applied across all operating points and platforms, provided their underlying components remain unaltered. This breaks away from conventional practices—such as tuning a generic overhead percentage—that overfit a single scenario and fail to generalize.
\end{enumerate}

Separating these sources is critical for accuracy. For instance, on a high-batch H200 decode step, throughput physics takes 12.7\,ms, but control and mechanism overheads add 11.1\,ms---comprising nearly half the total time, not just a rounding error. 

By adhering to component ownership, \sysname{} makes falsifiable predictions: removing a term primarily affects the serving conditions it represents (\S\ref{sec:eval-ablation}), parameters transfer safely across platforms (\S\ref{sec:eval-transfer}), and swapping a component requires re-calibrating only its specific owner (\S\ref{sec:eval-gates}).

\section{Evaluation}
\label{sec:eval}

\subsection{Experimental Setup}
\label{sec:eval-setup}

Our evaluation answers five research questions based on the predictions of our component-driven model:
\begin{itemize}
    \item \textbf{RQ1 (Attribution):} When we remove one overhead term, does accuracy degrade where we expect it to (\S\ref{sec:eval-ablation})?
    \item \textbf{RQ2 (Transferability):} Can the same parameters be reused across different hardware systems without re-calibration, and when only one component changes, is it sufficient to re-calibrate only the parameters associated with that component(\S\ref {sec:eval-transfer})?

    \item \textbf{RQ3 (Fidelity):} Does the simulation accurately reproduce both steady-state and transient serving dynamics (\S\ref{sec:eval-gates}, \S\ref{sec:eval-dynamics})?
    \item \textbf{RQ4 (Cost):} Is the simulator lightweight enough for extensive design sweeps (\S\ref{sec:eval-cost})?
    \item \textbf{RQ5 (Design):} Can high-fidelity simulation alter pre-silicon design decisions (\S\ref{sec:casestudy})?
\end{itemize}

\textbf{Configurations and Workloads.} Our primary evaluation uses DeepSeek-R1 (671B MoE, FP8) on 8$\times$H200 and H200/B200 disaggregated deployments (Dynamo with SGLang workers). For transient dynamics (\S\ref{sec:eval-dynamics}), we deploy Qwen3-32B on a 2$\times$H100 setup and validate against Prometheus/Grafana telemetry on LMBenchmark. To isolate pure pipeline overheads and prevent mixing KV transfers with collective communications (e.g., TP/EP), we strictly assign one GPU for prefill and one for decode, restricting NVLink traffic solely to KV transfers. To evaluate binding accuracy (\S\ref{sec:meanfield}), we analyze 44,220 iterations across 10 runs, encompassing steady-state decode, chunked prefill, and multi-turn contexts (up to 21K tokens).

\textbf{Validation Methodology.} To ensure an exact comparative validation, we execute unmodified benchmarks through our endpoint interface (\S\ref{sec:simserver}). We calibrate model parameters using a subset of the measurement points from the public InferenceX dashboard and evaluate the resulting model on the remaining held-out points. 

\subsection{Calibration Methodology}
\label{sec:validation}

\begin{table}[t]
\caption{\textbf{Application of Table~\ref{tab:ledger}'s transfer rules.}
Unlisted applicable families transfer unchanged.}
\label{tab:provenance}
\footnotesize
\centering
\setlength{\tabcolsep}{4pt}
\renewcommand{\arraystretch}{1.05}
\begin{tabular}{@{}p{0.16\columnwidth}p{0.17\columnwidth}p{0.30\columnwidth}p{0.27\columnwidth}@{}}
\toprule
Platform & Deployment & Change & Updated \\
\midrule
\rowcolor{black!8}
\multirow{2}{*}{H200}
& agg.\newline \textbf{(baseline)} & -- & -- \\
\addlinespace[3pt]
& disagg. & LLM engine build\newline Disagg. Framework
& \textbf{S, R} \\
\midrule
\multirow{2}{*}{B200}
& agg. & Blackwell GPU\newline LLM engine build
& \textbf{T, S} \\
\addlinespace[3pt]
& disagg. & +(B200 agg.)\newline Disagg. Framework\newline DPA, \mbox{P$\to$D handoff}
& \textbf{T, S, R} \\
\midrule
GB200\newline NVL72
& disagg. & +(B200 disagg.)\newline Grace host CPU\newline NVLink only
& \textbf{T, S, R} \\
\bottomrule
\end{tabular}
\end{table}

We calibrate component-specific overhead parameters from benchmark measurements and reuse them unchanged unless their owning components change.
For example, on the H200 baseline, host overheads are derived from request concurrency, KV costs from context length scaling, and collective costs from communication counts. Table~\ref{tab:ledger} defines the transfer rules for these parameters, and Table~\ref{tab:provenance} tracks their lifecycle across deployments. We initialize the baseline parameters on the aggregated H200 setup and following fix and resuse values rules when moving new deployment.

For instance, transitioning to an aggregated B200 updates only the hardware specifications and host control costs. Moving to a disaggregated architecture introduces new structural terms, such as prefill-to-decode pacing and data-parallel attention (DPA) residuals. When validating a disaggregated H200 on a different software snapshot, re-measuring the host control yielded values within $0.61$\,ms of the B200 baseline, confirming the reliability of this transfer method. This strict update policy prevents overfitting through arbitrary tuning. Any discrepancies between our simulated predictions and empirical data (e.g., anomalous compute efficiencies $>1.0$) indicate missing systemic factors, such as unmodeled framework optimizations or hidden execution overheads, which we then explicitly incorporate into the simulator.

\begin{figure}[t]
\centering
\includegraphics[width=\columnwidth]{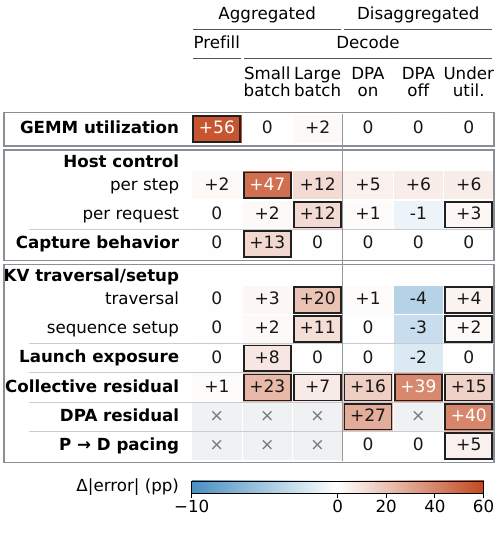}
\caption{\textbf{Ablation impact on absolute prediction error (pp).} Outlines mark the expected dominant effects of each term; $\times$ denotes absent mechanisms.}

\label{fig:ablation}
\end{figure}

\begin{figure*}[t]
\centering
\includegraphics[width=\textwidth]{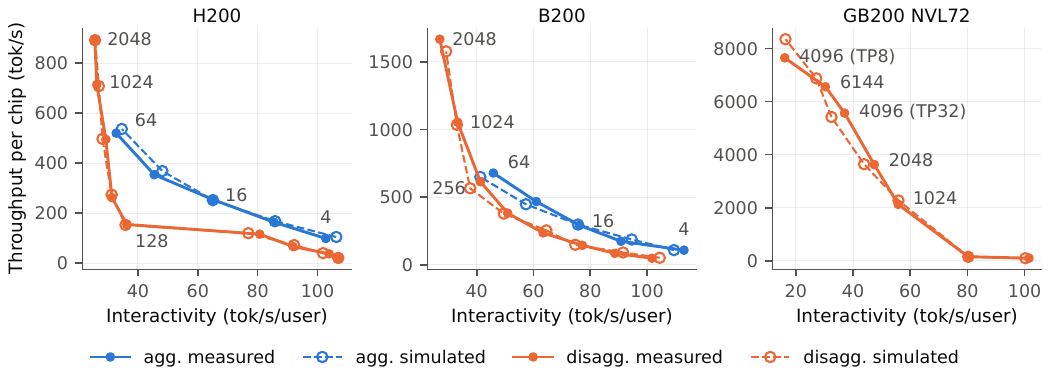}

\caption{\textbf{InferenceX throughput--interactivity frontiers.} Measured vs. simulated performance across H200, B200, and GB200 NVL72 (DeepSeek-R1 FP8, 1k/1k). Labels indicate concurrency.}
\label{fig:pareto}
\end{figure*}

\subsection{Sensitivity Analysis of Cost Terms}\label{sec:eval-ablation}

We conduct an ablation study to verify the necessity of the individual cost families in Table~\ref{tab:ledger}. By removing one term at a time, we evaluate whether prediction errors (reported as absolute percentage points, pp, in Figure~\ref{fig:ablation}) increase specifically under the serving conditions each term is designed to represent. Omitting the GEMM utilization term increases prefill prediction error by $56$\,pp, showing that peak compute rates alone do not capture effective GEMM performance. Per-step host control: Removing this increases small-batch decode error by $47$pp, demonstrating that CPU overhead remains significant even for small GPU batches. DPA synchronization: Removing this increases DPA-decode error by $40$pp. Other terms (e.g., KV traversal, collective costs) similarly dominate prediction errors under their corresponding conditions. These results justify maintaining separate cost terms rather than using a generic, interchangeable correction factor, an approach that conceptually aligns with LIMINAL~\cite{liminal}.

\subsection{Selective Transfer}
\label{sec:eval-transfer}

We evaluate parameter transferability across four dimensions, updating only the parameters associated with changed hardware or software components:
\begin{itemize}
    \item \textbf{Hardware (H200 $\to$ B200):} Updating only hardware specifications and host costs accurately predicts held-out data.
    \item \textbf{Stack (B200 $\to$ H200 disaggregated):} Reusing frozen B200 parameters on H200---re-measuring only host control---results in just a $3.1\%$ average error.
    \item \textbf{Workers (1 $\to$ 2 prefill):} The P→D pacing parameter models handoff control overhead separately from KV-data transfer time. We reuse it unchanged when scaling from one to two prefill workers within the same handoff-path class.
    \item \textbf{Platform (HGX $\to$ GB200 NVL72):} Re-calibrating only replaced components matches anchors within $2\%$. We avoid arbitrary overfitting, transparently reporting $7$--$12\%$ errors at extreme concurrencies where batching amortizes per-request overheads (e.g., state updates, KV cache allocation) to scale sublinearly ($\sim\!n^{0.6}$).

\end{itemize}

\subsection{End-to-End Steady-State Fidelity: InferenceX}
\label{sec:eval-gates}

We evaluate steady-state metrics and enforce the InferenceX dashboard's filtering criteria (e.g., pareto-optimal configurations). Furthermore, applying Little's law identifies the true bottleneck: a nominal "concurrency 2048" yields an effective pool of only $\approx$978 requests, proving the actual supply rate dictates the system load.

Figure~\ref{fig:pareto} validates our simulator across H200, B200, and GB200 environments. Across 27 operating points, the mean error is just 3.6\%, with all predictions falling within a 10\% margin. Robust generalization is evident, as held-out validation error (3.7\%) closely matches calibration error (3.5\%). Finally, evaluating a dense LLM (Qwen3-32B) yields a $3\times$ higher compute efficiency than the MoE baseline, exactly matching their arithmetic intensity ratio. This confirms that efficiency emerges from inherent architectural properties, not arbitrary tuning variables.

\begin{figure*}[t]
\centering
\includegraphics[width=\textwidth]{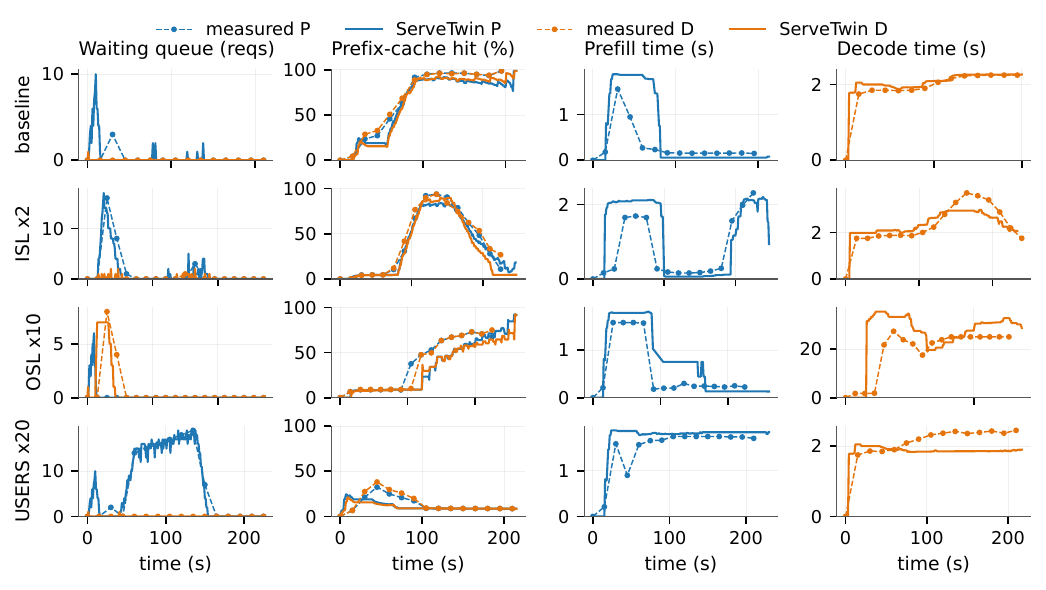}
\caption{\textbf{Transient serving dynamics on LMBenchmark.} Rows show the four scenarios; columns show waiting queues, prefix-cache hit rates, prefill time, and decode time. Blue and orange denote prefill and decode instances. Dashed markers show measurements and solid lines show simulation. Insets give measured and simulated timing summaries from Table~\ref{tab:fourway}.}
\label{fig:lmbench}
\end{figure*}

\subsection{Transient Serving Fidelity: LMBenchmark}
\label{sec:eval-dynamics}

To evaluate transient dynamics, we use LMBenchmark~\cite{lmbench} serving Qwen3-32B FP8 on two H100 GPUs (1 prefill, 1 decode worker) with vLLM and prefix caching. The baseline configures 1 user, 20 rounds, 2 QPS, 1K-token shared prompt, 10K-token history, and 100-token output. We evaluate three more variants: ISL$\times$2 (20K history), OSL$\times$10 (1K output), and USERS$\times$20 (20 users). 

Before the main run, 0.5\,s-interval two-round warm-ups and short requests populate reusable KV prefixes. During execution, users append to and resend accumulated conversations. Consequently, cache-hit rates naturally rise as repeated prefixes replace cold prefills. We update hardware specifications (memory capacity, HBM bandwidth, peak compute) to H100 and model inputs to Qwen3-32B at TP1. For the MoE-to-dense GEMM shift, we solely re-anchor the prefill GEMM utilization curve based on the baseline scenario. Despite shifting from SGLang to vLLM, we retain the H200 software overheads (KV-traversal, kernel launches, host costs, and CUDA-graphs).

Figure~\ref{fig:lmbench} compares simulations against actual measurements. ISL$\times$2 increases KV occupancy and cold-prefills, peaking the prefill queue at 16 requests. Initial reuse boosts hit rates, but growing contexts accelerate evictions, eventually lowering hit rates and forcing renewed prefills. Similarly, USERS$\times$20 distributes the 2 QPS across 20 separate per-user histories. These compete for the fixed KV pool without sharing prefixes, causing evictions before reuse. This sustains prefill workloads and lowers cache-hit rates. 

Conversely, OSL$\times$10 prolongs decoding, keeping KV blocks active. During warm-up, prefills accumulate while decode capacity is busy, causing a decode-admission spike. Longer responses delay subsequent multi-turns, causing cache-hit rates to rise slower than the baseline. However, high cache-hit rates ultimately reduce prefill work, though they cannot eliminate the extended decode phase.

\paragraph{Comparison with LLMServingSim.}
We compare our \sysname{} against LLMServingSim using its two provided trace modes: a replay mode with arrival timestamps exported from a specific benchmark, and a chained mode that triggers each turn only after its predecessor completes.\footnote{Release v1.1.0: the \texttt{sub\_requests} workload schema and its \texttt{tool\_duration} gap are present in the repository but are not described in its papers.} Both modes abort at the deployment's 33\,GiB KV budget (Figure~\ref{fig:state-failure}). To inspect their behavior beyond this failure, we relax only LLMServingSim's KV capacity to 4\,TB, while the measured deployment and \sysname{} retain the original budget. This diagnostic tests whether bypassing the capacity abort is sufficient to reproduce measured timing behavior.

Table~\ref{tab:fourway} shows that both two modes match baseline TTFT within $10\%$, but this agreement does not carry to the pressure scenarios. On ISL$\times$2 and USERS$\times$20, their TTFTs are $96$--$99.7\%$ below measurement, missing the large latency increase associated with prefill pressure. OSL$\times$10 is less uniform: replay underestimates TTFT by $42\%$, while chained adaptation overestimates it by $38\%$. Both modes overestimate ITL across all four scenarios, by $79$--$137\%$. Preserving turn dependencies alone therefore does not recover the measured workload-dependent timing behavior.

\subsection{Simulation Efficiency}
\label{sec:eval-cost}

Three forms of reuse cut trace-generation overhead. First, we generate each heavily repeated end-to-end trace expression once and reuse it. Second, rather than simulating individual layers, we exploit the layer-wise repetition of a Transformer to jump analytically to the end of all layer executions, a further $4.9\times$. Third, a shape cache groups slightly varying tensor shapes (e.g., batch sizes or sequence lengths) into size-based buckets that share one entry, instead of caching every shape separately. Consequently, all experiments, including full calibration sweeps, execute deterministically on a single commodity CPU in minutes.

\subsection{Pre-Silicon Design Exploration}
\label{sec:casestudy}

Motivated by the selective-transfer results in \S\ref{sec:eval-transfer}, we conduct two complementary design explorations. First, we investigate whether specific workloads benefit more from increased HBM bandwidth or expanded capacity. Second, under a fixed memory capacity, we evaluate whether bandwidth enhancements or scheduler and runtime optimizations better accommodate latency-sensitive versus throughput-oriented services. It is important to note that these are simulated sensitivity analyses, rather than equal-cost hardware comparisons or measurements of unreleased silicon.

\begin{table}[t]
\caption{\textbf{Simulation fidelity on LMBenchmark.} Post-warmup mean window medians; parentheses show signed errors vs. measurements. LLMServingSim uses relaxed capacity.}
\label{tab:fourway}
\footnotesize
\centering
\setlength{\tabcolsep}{2.2pt}
\begin{tabular}{@{}llrr@{}}
\toprule
Scenario & Source & \shortstack[r]{TTFT(s)} & \shortstack[r]{ITL(ms)} \\
\midrule
baseline & measured & 0.060 & 22.3 \\
\rowcolor{black!8}
         & \textbf{\sysname{} (calibration)} & \textbf{\boldmath 0.058 ($-$3\%)} & \textbf{22.5 (+1\%)} \\
         & LLMServingSim replay & 0.057 ($-$5\%) & 52.9 (+137\%) \\
         & LLMServingSim chained adapt. & 0.054 ($-$10\%) & 46.0 (+106\%) \\
\addlinespace[2pt]
ISL$\times$2 & measured & 3.14 & 29.7 \\
\rowcolor{black!8}
         & \textbf{\sysname{}} & \textbf{\boldmath 2.50 ($-$20\%)} & \textbf{\boldmath 26.8 ($-$10\%)} \\
         & LLMServingSim replay & 0.089 ($-$97\%) & 63.3 (+113\%) \\
         & LLMServingSim chained adapt. & 0.113 ($-$96\%) & 55.4 (+87\%) \\
\addlinespace[2pt]
OSL$\times$10 & measured & 0.219 & 25.0 \\
\rowcolor{black!8}
         & \textbf{\sysname{}} & \textbf{\boldmath 0.204 ($-$7\%)} & \textbf{28.5 (+14\%)} \\
         & LLMServingSim replay & 0.126 ($-$42\%) & 55.2 (+121\%) \\
         & LLMServingSim chained adapt. & 0.302 (+38\%) & 47.4 (+90\%) \\
\addlinespace[2pt]
USERS$\times$20 & measured & 15.2 & 24.0 \\
\rowcolor{black!8}
         & \textbf{\sysname{}} & \textbf{15.3 (+1\%)} & \textbf{\boldmath 18.6 ($-$23\%)} \\
         & LLMServingSim replay & 0.044 ($-$99.7\%) & 42.9 (+79\%) \\
         & LLMServingSim chained adapt. & 0.557 ($-$96\%) & 50.5 (+110\%) \\
\bottomrule
\end{tabular}
\end{table}

\paragraph{Bandwidth vs. Capacity.}

\begin{figure}[t]
\centering
\includegraphics[width=0.95\columnwidth]{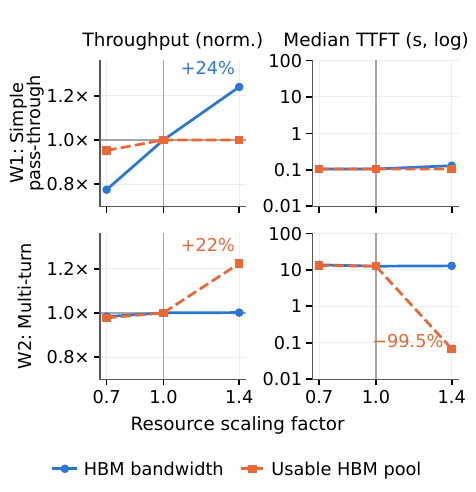}
\caption{\textbf{Workload-dependent HBM priorities.} Independent requests (top) and multi-turn sessions (bottom): normalized output throughput (left) and median TTFT on a shared logarithmic scale (right).}
\label{fig:sweep}
\end{figure}

Figure~\ref{fig:sweep} compares the sensitivity of two workloads to HBM bandwidth and capacity: independent single-turn requests (\textbf{W1}: 1K-input/1K-output, fixed concurrency of 64) and multi-turn sessions (\textbf{W2}: 20 users, 20 rounds per user, 11--13.2K-token prompts, and 100-token outputs). Both workloads use Qwen3-32B FP8 in a two-GPU H100 configuration with one prefill worker and one decode worker (1P1D). We separately scale HBM bandwidth and the usable memory pool to $\{0.7, 1, 1.4\}\times$ their baseline values, varying one while holding the other fixed. After subtracting the fixed model-weight footprint, 
the capacity sweep yields effective KV budgets of 14.1, 33.3, and 58.8\,GiB, respectively.

For \textbf{W1}, increasing bandwidth to $1.4\times$ baseline improves throughput by $24\%$, whereas increasing the memory pool provides no throughput benefit. Reducing the pool to $0.7\times$ baseline lowers throughput by $5\%$. These results are consistent with the baseline pool accommodating the active KV working set at the configured concurrency, while memory traffic during decoding remains a performance constraint. Although additional capacity can support larger batches when concurrency is increased, it provides no throughput gain under this fixed workload setting.

For \textbf{W2}, increasing bandwidth to $1.4\times$ baseline has little effect on end-to-end throughput. Increasing the memory pool to $1.4\times$ baseline, however, improves throughput by $22\%$ and reduces median TTFT from 12.55\,s to 0.068\,s. The prefix-cache hit rate rises from $22\%$ to $93\%$, while median queueing delay falls from 10.46\,s to approximately 0.01\,s. These changes are consistent with improved retention of reusable multi-turn contexts, reducing repeated prefill computation and queueing pressure. The limited throughput response to additional bandwidth suggests that accelerating memory-bound operations alone does not address the dominant serving constraints in this configuration.

Together, these workloads illustrate that the benefits of additional bandwidth and capacity depend on active KV demand, prefix reuse, and the stages that limit serving performance. In particular, capacity can improve responsiveness at a fixed user count by preserving reusable contexts across turns. Evaluating bandwidth--capacity trade-offs therefore requires accounting for cache residency and queueing dynamics alongside concurrency.

\begin{figure}[t]
\centering
\includegraphics[width=\columnwidth]{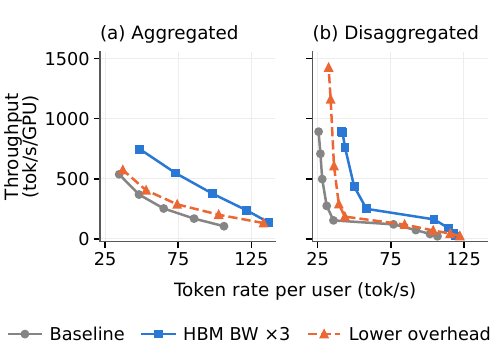}
\caption{\textbf{Bandwidth gains and overhead constraints.} (a) Aggregated and (b) disaggregated serving. ``Lower overhead'' denotes reductions in selected Scheduler and Runtime costs.}
\label{fig:service}
\end{figure}

\paragraph{Bandwidth vs. Scheduler and Runtime Costs.}
Figure~\ref{fig:service} evaluates hardware scaling versus software overhead reduction on a validated H200 setup (DeepSeek-R1 FP8, 1K-input/1K-output). We compare the baseline against tripling HBM bandwidth or reducing selected Scheduler costs (host control) and Runtime costs (prefill-to-decode (P$\to$D) pacing and DPA synchronization) by $40\%$. \textbf{Panel (a)} shows an aggregated TP8 configuration, while \textbf{Panel (b)} presents a disaggregated deployment scaling to extreme concurrencies.

Under low loads (e.g., concurrency 128), hardware bandwidth dominates. Because software queues and communication delays are practically non-existent at this stage, performance scales directly with memory bandwidth, increasing throughput by $28$--$63\%$. However, \textbf{Panel (b)} reveals a striking reversal under extreme loads. At a concurrency of 2048, tripling the bandwidth makes raw token generation $60\%$ faster, but overall throughput stalls completely, and median TTFT degrades to 52.96\,s. The hardware computes faster, but the software (scheduler and inter-worker synchronization) fails to manage the massive influx of requests, causing severe traffic congestion. Conversely, holding hardware fixed and cutting these Scheduler and Runtime costs by $40\%$ increases throughput by $60\%$ and halves TTFT to 22.92\,s. This indicates that at high concurrency, the binding constraint shifts from memory bandwidth to scheduler and runtime costs.
Crucially, the root cause of bottlenecks shifts depending not only on the workload characteristics but also on the current concurrency level being served.

\paragraph{Batch-Dependent Limits to Hardware Gains.}
\label{sec:discussion}

\begin{figure}[t]
\centering
\includegraphics[width=\columnwidth]{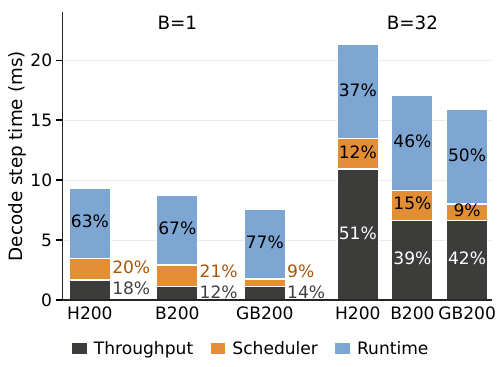}
\caption{\textbf{Batch-dependent decode cost decomposition.} Compares H200, B200, and GB200 at active batches $B=1$ and $B=32$ under a fixed DeepSeek-R1 FP8 setup (aggregated TP8, 1536-token FP8 KV cache). }
\label{fig:portion}
\end{figure}

Figure~\ref{fig:portion} presents a controlled decode-step comparison across different hardware platforms (H200, B200, GB200) to complement the sensitivity study. $B$ denotes the active batch size on a single decode worker, rather than system-wide concurrency. The total decode step time is decomposed into three model-attributed components: Throughput Cost (analytical trace execution), Scheduler Cost (exposed host CPU work), and Runtime Cost (KV traversal, sequence setup, kernel launch, and collective setup residuals). This comparison uses no capture padding, so the Scheduler component contains only exposed host work. 

We utilize the hardware calibration profiles from \S\ref{sec:eval-transfer}. H200 and B200 share identical runtime coefficients, isolating the pure effect of their hardware specification differences. For GB200, the per-step Scheduler cost (host control) is explicitly reduced from 1.8\,ms to 0.635\,ms. Consequently, the GB200 results represent a transferred platform-model prediction rather than an empirical measurement.

This breakdown explains why hardware upgrades fail to deliver proportional performance gains at small batch sizes. At a low batch size ($B=1$), Runtime Cost dominates, accounting for 63--77\% of the total step time. Upgrading from H200 to B200 reduces the Throughput Cost from 1.67\,ms to 1.09\,ms ($1.53\times$ faster), but the total step time only marginally improves from 9.34\,ms to 8.76\,ms ($1.07\times$). 

Conversely, at a high batch size ($B=32$), Throughput Cost accounts for a significantly larger share. Moving from H200 to B200 reduces this cost from 10.94\,ms to 6.64\,ms ($1.65\times$ faster), resulting in a more substantial total step time reduction from 21.34\,ms to 17.04\,ms ($1.25\times$).

\section{Related Work}
\label{sec:related}

\textbf{Simulators and Analytical Models.} Beyond the tools discussed in \S\ref{sec:existing}, most simulators (e.g., APEX~\cite{apex}) are validated solely on custom internal testbeds. LIMINAL~\cite{liminal} bounds decode performance using a single lumped overhead term. In contrast, \sysname{} isolates specific mechanism costs to accurately model state-dependent queueing behaviors.

\textbf{Overhead Characterization.} While KernelSight-LM~\cite{kernelsight} profiles host-side overheads via direct instrumentation, \sysname{} complements this approach by analytically inferring these overheads from public measurements and verifying them through component-level parameter transfer.

\textbf{Disaggregated Systems.} \sysname{} directly models the underlying mechanisms of modern disaggregated serving architectures pioneered by Splitwise, DistServe, and Mooncake.

\section{Conclusion}
\label{sec:conclusion}

\sysname{} predicts LLM serving performance by integrating a KV-lifecycle state machine with a specification-driven analytical backend. Validated against independent public benchmarks, it achieves a $3.6\%$ mean error on InferenceX frontiers and reproduces queue and cache-reuse dynamics in multi-turn scenarios. By updating calibration parameters strictly based on architectural changes, \sysname{} successfully transfers to new platforms, predicting GB200 NVL72 performance within a $2\%$ margin. This capability allows architects to generate serving curves directly from pre-silicon specifications. Finally, our design studies demonstrate that hardware upgrade priorities must align with workload states and service targets.

\bibliographystyle{ACM-Reference-Format}
\bibliography{refs}

\end{document}